\documentclass[12pt]{article}

\usepackage{amsmath}
\begin{document}

\title{The Influence of Quantum Field Fluctuations\\
 on Chaotic Dynamics of Yang--Mills System}

\author{V.~I. KUVSHINOV and A.~V. KUZMIN \\
Institute of Physics, Scorina av. 68, 220072, Minsk, Belarus \\
E-mails: kuvshino@dragon.bas-net.by, avkuzmin@dragon.bas-net.by}

\maketitle

\begin{abstract}
\noindent On example of the model field system we demonstrate that quantum
fluctuations of non-abelian gauge fields leading to radiative corrections to Higgs
potential and spontaneous symmetry breaking can generate order region in phase space
of inherently chaotic classical field system. We demonstrate on the example of another
model field system that quantum fluctuations do not influence on the chaotic dynamics
of non-abelian Yang--Mills fields if the ratio of bare coupling constants of
Yang--Mills and Higgs fields is larger then some critical value. This critical value
is estimated.
\end{abstract}

\strut\hfill

\noindent A steady interest to chaos in gauge field theories (GFT) \cite{BookGFT} is
connected with the facts that all four fundamental particle interactions have chaotic
solutions \cite{4}. There are a lot of footprints on chaos in HEP \cite{Kawabe, N},
nuclear physics (energy spacing distributions) \cite{Nuclear, Bunakov}, quantum
mechanics (QM) \cite{QM}.

Originally phenomenon of chaos was associated with problems of classical mechanics and
statistical physics. Substantiation of statistical mechanics initiated intensive study
of chaos and uncovered its basic properties mainly in classical
mechanics~\cite{Krylov}.
 One of the main results in this direction was a creation of KAM theory
and understanding of phase space structure of Hamiltonian systems. It was clarified
that the root of chaos is local instability of dynamical system. Local instability
leads to mixing of trajectories in phase space and thus to non-regular behavior of
system and chaos (for review see~\cite{Lihtenberg, Zaslavsky}).

Large progress is achieved in understanding of chaos in semi-classical regime of
quantum mechanics~\cite{B, Robnic}.

Investigation of stability of classical field solutions faces difficulties caused by
infinite number of degrees of freedom. That is why authors often restrict their
consideration by the investigation of some model field configurations~\cite{Kawabe,
SHS, regular}.

There are papers devoted to chaos in quantum field theory~\cite{Q2,PLA}. Nevertheless
there is no generally recognized definition of chaos for quantum systems in QM (beyond
semiclassical approximation) and QFT~\cite{Bunakov}. This fact restricts use of chaos
theory in the field of elementary particle physics.

It was analytically \cite{Savvidy, Salasnich, we3} and numerically \cite{Kawabe,
regular} shown that classical gauge Yang--Mills theories (GYMT) are inherently chaotic
theories. Particulary, it was shown for spatially homogeneous field
configurations~\cite{SHS} that spontaneous symmetry breakdown leads to appearance of
order-chaos transition with rise of density of energy of classical gauge fields
\cite{regular, Salasnich, we3}, whereas
  dynamics of gauge fields in the absence of spontaneous symmetry breakdown is
  chaotic at any
 density of energy~\cite{Savvidy}. This conclusion was supported
 by studying the stability of topological solutions~\cite{Kawabe}. Study
 of chaos
 in classical GYMTs revealed several new problems. One of them is to understand
 the role of Higgs fields from the viewpoint of their influence on
  chaotic dynamics of classical gauge fields. It
 was demonstrated that classical Higgs fields regularize
  chaotic dynamics of classical gauge fields at low densities of energy and lead to appearance
 of order-chaos transition~\cite{Kawabe, regular}. The most of the results concerning
 chaos in GYMTs are obtained in classical theories. The question about
 chaotic properties of quantum GYMTs remains open. However, it was demonstrated that
 quantum fluctuations of abelian gauge field leading to spontaneous symmetry breaking
 via Coleman--Weinberg effect~\cite{effpot} regularize chaotic dynamics of spatially
 homogeneous system of Yang--Mills and Higgs fields at small densities of energy~\cite{1997}.

 In this Letter we demonstrate that the same effect takes place in the case of
 non-abelian gauge field theory such as the theory of electroweak interactions.
 Namely, the ``switching on'' of quantum fluctuations
 of vector gauge fields leads to ordering at low densities of energy,
 order-to-chaos transition with the rise of density of energy of gauge fields
 occurs (compare with~\cite{1997}).
 This phenomenon escapes one's notion under classical consideration and
 thus we make the step to understanding the role of chaos in GYMTs. Also
 we note that if the ratio of the coupling constants of Yang--Mills and Higgs fields is
 larger than some critical value then quantum corrections do not affect the chaotic
 dynamics of gauge and Higgs fields.

 Consider $SU(2)\otimes U(1)$ gauge field theory with real massless scalar
 field $\rho$ with the Lagrangian
\begin{gather}
 L=-\frac{1}{4}G_{\mu \nu }^{a}G^{a}{}^{\mu \nu }-\frac{1}{4}H_{\mu \nu
}H^{\mu \nu }+\frac{1}{8}g^{2}\rho ^{2}\left( W_{1}^{2}+W_{2}^{2}+\frac{Z^{2}%
}{\cos ^{2}{\theta _{w}}}\right) \nonumber\\
\phantom{L={}}{}+\frac{1}{2}\partial _{\mu }\rho \partial ^{\mu }\rho -
\frac{1}{4!}\lambda \rho ^{4}.\label{Lagrangian}
\end{gather}
Here $\lambda $ denotes a self-coupling constant of scalar field, $g$~---
self-coupling constant of non-abelian gauge fields, $\theta _{w}$ is Weinberg angle,
$A_{\mu }$ corresponds electro-magnetic field, $W_{\mu }^{1}$, $W_{\mu }^{2}$ describe
W-bosons and $Z_{\mu }$~--- neutral Z-boson.

We used the following denotations
\begin{gather*}
G^{a}_{\mu \nu}=\partial_{\mu}W^{a}_{\nu} -
\partial_{\nu}W^{a}_{\mu} +
g\varepsilon^{abc}W^{b}_{\mu}W^{c}_{\nu} , \qquad a=1,2,3;\\
H_{\mu \nu} = \partial_{\mu}W^{0}_{\nu} -
\partial_{\nu}W^{0}_{\mu}.
\end{gather*}
To study the dynamics of classical gauge fields from the viewpoint of chaos we use
spatially homogeneous solutions~\cite{SHS}. Their dynamics in the classical vacuum of
scalar field $\rho =0$ is chaotic at any densities of energy~\cite{Savvidy}.

Situation qualitatively changes if we take into account quantum fluctuations of vector
fields. It is caused by the known fact that the state $\langle \rho\rangle = 0$ is not
a vacuum state in this case. Here $\langle \rho\rangle $ denotes vacuum quantum
expectation value of scalar field related with its classical vacuum value $\rho$ as
follows $\langle \rho \rangle= \rho + \mbox{quantum corrections}$. To find a true
vacuum of scalar field in this case we use the method of the effective
potential~\cite{effpot}.

One loop  effective potential generated by the Lagrangian (\ref{Lagrangian}) has the
form (see also~\cite{Huang})
\begin{gather}
 U\left( \langle \rho\rangle \right)
= \frac{1}{4!}\lambda \langle\rho \rangle^{4} + \frac{3g^{4}\langle \rho
\rangle^{4}}{128\pi ^{2}}\left( -\frac{1}{2}+\ln {\frac{
g^{2}\langle \rho\rangle^{2}}{2\mu ^{2}}}\right)  \nonumber \\
\phantom{U\left( \langle \rho\rangle \right)=}{} +\frac{3g^{4}\langle \rho\rangle
^{4}}{256\pi ^{2}\cos ^{4}{\theta _{w}}}\left( -\frac{1}{2} +\ln {\frac{g^{2}\langle
\rho \rangle^{2}}{2\mu ^{2}\cos ^{2}{\theta _{w}}}}\right).\label{eff}
\end{gather}
Where $\mu^{2}$ is a renormalization constant. Here we took into account contributions
of all Feynman diagrams with one loop of any ($W_{1}$, $W_{2}$, $Z$ or $A$) gauge
field and external lines of Higgs field. This potential leads to spontaneous symmetry
breaking and non-zero vacuum expectation value of scalar $\rho$-field appears.
Classical vacuum of scalar field is $\rho =0$, but it is not so in quantum case,
because of Coleman--Weinberg effect~\cite{effpot}.

In the case of the Lagrangian (\ref{Lagrangian}) it is known that vacuum state of the
scalar field becomes degenerated and $\langle \rho \rangle \neq 0$ if we take into
account quantum properties of gauge fields~\cite{Huang}.

It is easy to calculate that the squared  vacuum expectation value equals
\begin{equation*}
\langle \rho\rangle_{v}^{2} = \frac{2 \mu^{2}}{g^{2}} \exp\left[ \frac{\left( 18g^{4}
\ln{\cos{\theta_{w}}} - 32\pi^{2} \lambda \cos^{4}{\theta_{w}} \right)}{9g^{4}\left( 1
+ 2\cos^{4}{\theta_{w}} \right)} \right].
\end{equation*}
Its existence qualitatively changes chaotic behavior of spatially homogeneous
solutions in the quantum vacuum of scalar field compared to pure classical
consideration. Order-to-chaos transition occurs with the rise of density of energy of
non-abelian gauge fields (see also~\cite{1997}).

To demonstrate this in more evident form let us consider a simplified model system.
Full consideration does not add any new physical content.

We will investigate fields of the following form
\begin{equation}\label{conf}
W_{i}^{a}=e_{i}^{a}q^{a}\left( \tau \right),\qquad a=1,2, \quad i=1,2,3; \qquad
\left(\vec{e^{a}}\right)^{2}=1 , \qquad \vec{e^{1}}\vec{e^{2}}=\cos{\xi}.
\end{equation}
Here $e^{a}_{i}$ are constant vectors. Other {\it classical} gauge fields for
simplicity are put to be equal to zero, but their quantum fluctuations are included
and give contribution to the potential~(\ref{eff}).

Lagrangian describing dynamics of the model field system has the following form
\begin{equation}  \label{eq:smallL}
H = \frac{1}{2}\left(p_1^2 + p_2^2 \right) + \frac{1}{8} g^{2}\langle
\rho\rangle_{v}^{2} \left(q_{1}^{2} + q_{2}^{2}\right) + \frac{1}{2}
g^{2}q_{1}^{2}q_{2}^{2}\sin^{2}{\xi}.
\end{equation}
Here $p_1$, $p_2$ are momenta of gauge fields, $0<\sin^{2}{\xi}<1$ is a free
parameter. Influence of quantum fluctuations on the classical model field
configuration (\ref{conf}) is taken into account in~(\ref{eq:smallL}). Lagrangian of
the classical (no quantum corrections) model system follows from~(\ref{eq:smallL}) if
one puts $\langle\rho\rangle_{v} = 0$.

To analyze dynamics of the system from the viewpoint of chaos we use well known
technique~\cite{Zaslavsky, Rep}. Particulary, we reduce equations of motion expressed
in action-angle variables (for instance see \cite{Lihtenberg}) to discrete mapping by
angle variable $\overline{\psi}$ and calculate parameter of local instability
$K=|\delta \overline{\psi}_{n+1}/\delta \overline{\psi}_{n} - 1|$ defined
in~\cite{Rep} ($K>1$~--- motion is locally unstable and chaotic, $K\leq 1$~--- motion
is stable). Here $\overline{\psi}_{n+1}$ and $\overline{\psi}_{n}$ denote values of
angle variable at $(n+1)$-th and $n$-th time step respectively. Form of the map and
detailed calculations can be found in~\cite{we3} where model system~(\ref{eq:smallL})
was considered in different physical context.

Let us put $\langle \rho\rangle_{v}=0$ (that corresponds to absence of gauge fields
quantum fluctuations). In this case, as it was clarified in~\cite{Savvidy}, dynamics
of spatially homogeneous field configurations in the classical vacuum of real scalar
field is always chaotic, at any, even small densities of energy. This conclusion
agrees with the results of~\cite{regular, Salasnich, we3}.

If vector gauge fields quantum fluctuations are ``switched on'', and therefore
$\langle \rho\rangle_{v} \neq 0$, then parameter $K$ can be calculated (see
\cite{we3}) and equals
\begin{equation}\label{K}
  K=\frac{8E\sin^{2}{\xi}}{g^{2}\langle \rho\rangle_{v}^{4}}.
\end{equation}
Here $E$ is the density of energy of the spatially homogeneous model field
system~(\ref{eq:smallL}). From~(\ref{K}) it follows that at small densities of energy
motion is stable ($K<1$). At large enough densities of energy motion becomes unstable
and non-regular ($K>1$). Thus if the density of energy of the model field system
increases, we obtain order-to-chaos transition. Critical density of energy $E_{c}$
corresponding to this transition is given by the following expression
\begin{equation*} 
E_{c}=\frac{1}{8}\frac{g^{2}\langle \rho\rangle_{v}^{4}}{\sin^{2}{\xi}}.
\end{equation*}
Thus at the densities of energy less than $E_{c}$ dynamics of the field system is
regular and at the densities of energy larger than $E_{c}$ we obtain regions of
unstable motion and chaos in the phase space of the system.

We can conclude that quantum properties of gauge fields are essential since they
quali\-tatively change chaotic dynamics of classical gauge field configurations.
Quantum gauge field fluctuations generated by the Lagrangian (\ref{Lagrangian}) lead
classical non-abelian Yang--Mills fields to order at low densities of energy due to
the same mechanism as it was demonstrated for abelian one~\cite{1997}. Thus
order-to-chaos transition missed under the classical consideration can be obtained if
one includes quantum effects.

Similar analysis can be made for boson sector of electro-weak theory including Higgs
potential. In this case quantum corrections are considered against the background of
the classical vacuum value of Higgs fields. Thus quantum fluctuations of fields lead
only to quantitative changes.

Now in contrary we demonstrate that under the certain conditions quantum corrections
to the Higgs potential leading to spontaneous symmetry breaking practically do not
change chaotic behavior of classical Yang--Mills and Higgs fields. In order to affect
essentially their dynamics the ratio $\lambda / g^{4}$ has to be less then some
critical value which is calculated.

Starting from the Lagrangian (\ref{Lagrangian}) we build new model field system
including Yang--Mills and Higgs fields. We consider spatially homogeneous fields. For
non-abelian gauge fields we use the same assumptions and denotations as
 in~(\ref{eq:smallL}). In order to simplify the problem we put $\xi =0$ which means that
the fields of $W^+$ and $W^-$ gauge bosons have the same linear polarization.
Therefore the Hamiltonian of the model field system has the form
\begin{equation}\label{Hamiltonian}
H = \frac{1}{2}\left(p_1^2 + p_2^2 + p^2\right) + \frac{1}{8} g^{2} \langle
\rho\rangle^{2} \left(q_1^2 + q_2^2\right) + U(\langle \rho\rangle).
\end{equation}
Here $p_1$ and $p_2$ are momenta of gauge fields and $p$ is a momentum of Higgs field.
Now we make the substitution $q_1 = r \cos{\varphi}$ and $q_2 = r \sin{\varphi}$. It
is easy to see that $\varphi$ is a~cyclical variable and therefore its conjugate
momentum $p_\varphi = r^2 \dot{\varphi}$ is a constant of motion.
Hamiltonian~(\ref{Hamiltonian}) can be written is the form
\begin{equation}\label{H}
  H = \frac{1}{2} \left(p_r^2 + p^2\right) + \frac{1}{8} g^{2} \langle \rho\rangle^2 r^2 +
U(\langle\rho\rangle).
\end{equation}
Here $p_r$ is a conjugate momentum of $r$. We neglected by the term $p_\varphi^2 /
r^2$ compared to the term proportional to $r^2$ considering further classical
Yang--Mills fields with high intensity. Using well known technique based on Toda
criterion of local instability~\cite{Salasnich, PLA} one can obtain that there is an
order to chaos transition with the rise of the density of energy of the system.
Critical density of energy (minus vacuum density of energy which is non-zero) is given
by the following relation
\begin{equation}\label{NewE}
 E_c = \frac{3 \mu^4}{64 \pi^2} \exp{\left( 2 \alpha_w - \frac{2\lambda}{g^4}
  \beta_w \right)}\left(1 + \frac{1}{2 \cos^4{\theta_w}}\right) \left[ 1 - 7 e^{-48 \Lambda_w \beta_w}  \right].
\end{equation}
Here the following denotations are used
\begin{gather*}
\alpha_w = \frac{2 \ln{\cos{\theta_w}}}{1 + 2 \cos^{4}{\theta_w}}, \qquad
\beta_w = \frac{32 \pi^2 \cos^{4}{\theta_w}}{9\left(1 + 2 \cos^{4}{\theta_w}\right)}, \\
\Lambda_w = \frac{3}{128 \pi^2}\left( 1+ \frac{1}{2\cos^{4}{\theta_w}} \right).
\end{gather*}
If one does not take into account quantum corrections to Higgs potential it can be
shown that the behavior of the system~(\ref{H}) is chaotic at any density of energy.
From the expression~(\ref{NewE}) it is seen that $E_c$ can be exponentially suppressed
if the ratio of bare coupling constants of Higgs and gauge fields is larger then some
critical value. If this ratio is large enough $E_c$ exponentially close to zero and
one does not observe any order to chaos transition. Otherwise, if the ratio of
coupling constants is small enough, the critical density of energy given by the
expression~(\ref{NewE}) have detectible value. Thus the magnitude of the critical
density of energy strongly depends on the value of the ratio $\lambda / g^4$. Namely,
if the following condition is true
\begin{equation*}
\frac{\lambda}{g^4} < \frac{2}{\beta_w},
\end{equation*}
then quantum radiative corrections to Higgs potential affect the chaotic dynamics of
the model system. Otherwise, at least one-loop effective potential can not essentially
regularize chaotic dynamics.

In conclusion, we demonstrated that quantum fluctuations of non-abelian gauge fields
leading to effective potential of Higgs field can regularize chaotic dynamics of
Yang--Mills fields at low densities of energy. We showed also that if Higgs field is
considered as a~dynamical variable then under the conditions stated above quantum
corrections do not affect classical dynamics of gauge and Higgs fields.

Quantum corrections are known to be important in the framework of QCD, where
spontaneous symmetry breakdown is suggested to appear via Coleman--Weinberg effect.
Therefore we can expect that there is order-to-chaos transition in QCD at low
densities of energy as it is described above.

It is known that classical chaos influences on quantum properties of the system, for
instance, on the rate of quantum tunnelling (chaos assisted tunnelling), see~\cite{B}
and references therein. Here we have demonstrated that inverse effect exists also.
Quantum fluctuations of the system can change its classical chaotic behavior.

\label{kuvshinov-lastpage}


\begin{thebibliography}{99}
\small

\bibitem{SHS} Baseyan G~Z, Matinyan S~G and  Savvidy G~K, Nonlinear Plane Waves in
Massless Yang--Mills Theory, {\it Pis'ma Zh. Eksp. Teor. Fiz.} {\bf 29} (1979),
641--644.

\bibitem{regular} Berman G~P, Mankov Yu~I and  Sadreyev A~F, Stochastic Instability of Classical
Homogeneous SU(2)$\, \otimes\,$U(1) Fields with Spontaneous Broken Symmetry, {\it Zh.
Eksp. Teor. Fiz. (JETP)} {\bf 88} (1985), 705--714.

\bibitem{BookGFT} Biro T~S, Matinyan S~G and Muller B, Chaos and
Gauge Field Theory, World Scientific, 1994.

\bibitem{Q2} Biro T S, Muller B and Matinyan S G, Chaotic Quantization of Classical
Gauge Fields, hep-th/0010134.

\bibitem{B} Bohigas O, Tomsovic S and Ullmo D,
Manifestations of Classical Phase Space Structures in Quantum Mechanics, {\it Phys.
Rep.} {\bf 223} (1993), 43--133.

\bibitem{Nuclear} Bohr A and Mottelson B, Nuclear Structure, Benjamin, New York, 1967.

\bibitem{Bunakov}  Bunakov V~E, Chaos, Quantum Chaos and Nuclear Physics,
in Proceedings of the XXXII Winter School of PNPI,  PNPI Press, S.-Petersburg,1998,
5--33.

\bibitem{effpot}  Coleman  S and  Weinberg E, Radiative Corrections as the Origin of
Spontaneous Symmetry Breaking, {\it Phys.Rev.} {\bf D7} (1973), 1888--1910;\\
Lee B~W and Zinn-Justin J, Spontaneously Broken Gauge Symmetries. IV. General Gauge
Formulation, {\it Phys. Rev.} {\bf D7} (1973), 1049--1056.

\bibitem{4} Deterministic Chaos in General Relativity, Editors:
Hobill et al., NATO ASI Series B, Physics,  Vol.~332, Plenum Press, New York, 1994.

\bibitem{QM}
Gutzwiller M~C, Chaos in Classical and Quantum Mechanics, Springer, New York, 1990;\\
Chaos and Quantum Physics, Editors: Giannoni M-J, Voros~A and Zinn-Justin~J,
North-Holland, Amsterdam, 1991;\\
Brack M and Bhaduri R~K, Semiclassical Physics, Addison-Wesley, 1997.

\bibitem{Huang}  Huang K,  Quarks, Leptons and Gauge Fields, World
                Scientific, 1982, Chapter 10.

\bibitem{Kawabe} Kawabe T and Ohta S, Onset of Chaos in Time-Dependent Spherically
Symmetric SU(2) Yang--Mills Theory, {\it Phys. Rev.} {\bf D41} (1990), 1983--1988;\\
Kawabe T and Ohta S, Order-to-Chaos Transition in SU(2) Yang--Mills--Higgs Theory,
{\it
Phys. Rev.} {\bf D44} (1991), 1274--1279;\\
 Kawabe T, Onset of Chaos in SU(2) Yang--Mills--Higgs Theory with Monopole, {\it Phys.
Lett.} {\bf B274} (1992), 399--403;\\
 Kawabe T and Ohta S, Onset of Chaos in Yang--Mills--Higgs Systems, {\it Phys. Lett.}
{\bf B334} (1994), 127--131;\\
Kawabe T, Deterministic Chaos in Spatially Homogeneous Model of Abelian--Higgs Theory,
{\it Phys. Lett.} {\bf B343} (1995), 254--258.

\bibitem{Krylov} Krilov N~S, Papers on Statistical Physics Foundation, Akad. Nauk
SSSR, Moskow--Leningrad, 1950.

\bibitem{we3} Kuvshinov V~I and  Kuzmin A~V, Order-to-Chaos Transition in SU(2)
Spatially Homogeneous Model Field System, {\it Nonl. Phenom. in Comp. Sys.} {\bf 4}
(2001), 64--66.

\bibitem{PLA} Kuvshinov V~I and  Kuzmin A~V, Towards Chaos Criterion in Quantum Field
Theory, {\it Phys. Lett.} {\bf A296} (2002), 82--86.

\bibitem{Lihtenberg} Lichtenberg A~J and Lieberman M~A, Regular and Stochastic Motion,
Springer, Berlin, 1983.

\bibitem{1997} Matinyan S~G and Muller B, Quantum Fluctuations and Dynamical Chaos,
{\it Phys. Rev. Lett.} {\bf 78} (1997), 2515--2518.

\bibitem{N} Nielsen H~B, Rugh H~H and Rugh S~E, Chaos and Scaling in Classical
Non-Abelian Gauge Fields, chao-dyn/9605013.

\bibitem{Robnic}
Robnik M, Aspects of Quantum Chaos in Generic Systems, {\it Open Sys. and Information
Dyn.} {\bf 4} (1997), 211--240.

\bibitem{Salasnich} Salasnich L, Quantum Signature of the Chaos-Order Transition in a
Homogeneous SU(2) Yang--Mills--Higgs System, nucl-th/9707035.

\bibitem{Savvidy} Savvidy G~K, The Yang--Mills Classical Mechanics as a Kolmogorov
K-system, {\it Phys. Lett.}  {\bf B130} (1983), 303--307.

\bibitem{Zaslavsky}
Zaslavsky G~M, Stochasticity of Dynamical Systems, Nauka, Moscow, 1984;\\
Sagdeev R~Z and Zaslavsky G~M, Introduction in Nonlinear Physics, Nauka, Moscow, 1988.

\bibitem{Rep} Zaslavsky G~M, Stochasticity in Quantum Systems, {\it Phys. Rep.} {\bf 80}
(1981), 159--250.
\end{thebibliography}
\end{document}